\documentclass[11pt,twoside]{article}
\usepackage{asp2010}

\resetcounters

\begin{document}

\title{Protoplanetary Disks of Binary Systems in Orion}
\author{S. Daemgen$^1$, M.~G. Petr-Gotzens$^1$, and S. Correia$^2$}
\affil{$^1$European Southern Observatory, Karl-Schwarzschildstr. 2, 85748 Garching, Germany}
\affil{$^2$Astrophysikalisches Institut Potsdam, An der Sternwarte 16, 14482 Potsdam, Germany}

\begin{abstract}
Dusty primordial disks surrounding young low-mass stars are revealing tracers of stellar and planetary formation. The evolution and lifetime of these disks define the boundary conditions of the mechanisms of planet formation. Stellar companions, however, can significantly change this evolution through their tidal interactions. Stellar evolution and planet formation in binaries have to respond to an environment of truncated, quickly disappearing disks---very different compared to an isolated star environment.

In order to investigate details of the influence of binarity on circumstellar disk evolution, we obtained adaptive optics supported near-infrared imaging and spectroscopy of the individual components of 22 low-mass binaries in the well-known Orion Nebula Cluster. Brackett gamma emission, which we detect in several systems, is used as a tracer for the presence of an active accretion disk around each binary component.

We find a low fraction of accreting binary components, when compared to the disk fraction of single stars in the ONC. This might indicate a significantly faster evolution of disks in binaries. This finding is paticularly interesting, since the target sample consists of wide $>$\,100\,AU binaries---separations for which the disks are typically expected to evolve similarly to single star disks.
\end{abstract}

\section{Introduction}
A wealth of studies have explored the evolution of primordial circumstellar disks in star forming regions. These mostly focused on single systems, since they are easier to observe and, from a theoretical point of view, less complex than multiple systems and hence easier to model. However, 42\% of all late-type field stars are bound in binary systems \citep{fis92} and the binary frequency is even higher for stars of solar mass and above \citep{duq91}. Assuming that most stars are formed in clusters these numbers are lower limits for the initial binary frequency shortly after formation, because tidal interaction in the cluster will disrupt many systems. It can hence be assumed that binaries are the most important branch of star formation.

Disk evolution in binaries is controversially discussed in the literature. The total mass of a disk is likely reduced as long as the binary separation is less than $\sim$3 times the typical disk size in the star forming region in focus \citep{arm99}. This seems to agree with observational studies \citep[e.g.][]{bou06,mon07,cie09} that show that the frequency of disks in binary systems is significantly lower than that in single stars for systems separated by 100\,AU and less---a hint at an overall faster evolution of circumstellar disks in binaries, since, in addition to the dynamically removed outer parts, the innermost regions as measured by means of accretion and hot dust are also missing. Other studies like the mid-IR observations of \citet{pas08}, however, detect no difference in the evolution of the disk between singles and binaries.

Primordial disks are not only indicators for the formation of the star itself, but also contain the material for the formation of planetary systems. If disks in binaries evolve differently from those in single stars, then the properties of the population of planets found in binaries will likely reflect those differences. Today, more than 43 planets are known to orbit one of the components of a binary system \citep{mug09}, most of which are separated by 30\,AU and more. Interestingly, systems with both components orbited by their own planet are disproportionally rarely observed---to the knowledge of the authors no system has been published so far. Although not entirely free from observational selection effects, this suggests a differential evolution of the individual disks of a binary system. Investigating the presence of disks with respect to their appearance around the higher and lower mass components of a large number of binaries will hence help to understand the appearance of planets in binaries and restrict the planet formation process in terms of available formation time.

\section{Methods \& Goals of the Project}
This contribution describes a study of the Orion Nebula Cluster (ONC) investigating a sample of 22 binaries for signs of accretion and dust disk presence around each separate component. Our observations of binaries in the ONC will address, among others, the following questions:
\begin{itemize}
  \item What is the frequency of circumstellar disks in primaries and secondaries of young binaries?
  \item Do disks around secondaries disappear sooner than around primaries?
  \item How do disks evolve in the presence of a stellar companion?
\end{itemize}

In order to approach the answers to above questions we are using photometric as well as spectral information of the individual components of all target binaries. $JHK$ photometry allows us to assess near infrared excess emission and hence the presence of warm dust in an optically thick inner disk. Spectroscopy in $K$-band enables us to determine spectral types as well as to identify the actively accreting binary components of the sample. The latter is achieved through measuring the strength of Brackett gamma (Br$\gamma$) emission produced when accreting from the inner disk.

\section{Targets \& Observations}
\subsection{Sample Selection}
The Orion Nebula Cluster is one of the nearest \citep[414$\pm$7\,pc;][]{men07}, young star cluster. It is $\lesssim$1($\pm2$)\,Myr in age \citep{hil97} and it has been intensively investigated for its circumstellar disks \citep[e.g.][]{hil98,da_10} and its binary content \citep{sim99,khl06,rei07}. The ONC hence features excellent conditions to study the evolution of circumstellar disks in binaries and allows us to compare our findings to results from single stars.

We selected 22 visual binaries in the ONC. The observed projected separations range from 0.25 to 1.1\,arcsec, which corresponds to roughly 100--400\,AU at the distance of the ONC. Magnitude differences of the binary components range from 0.1 to $\sim$3\,mag in $H$ and $K$-band. Since all of the targets are likely to be members of the ONC \citep{rei07} and the separations are small, we can assume that the targeted binaries are gravitationally bound companions. Membership and hence physical binarity is further supported by analysis of the observed spectra and photometry, which suggest late spectral types at moderate luminosity ruling out background giants.

\subsection{NIR Photometry \& Spectroscopy}
\begin{figure}[t]
\plotone{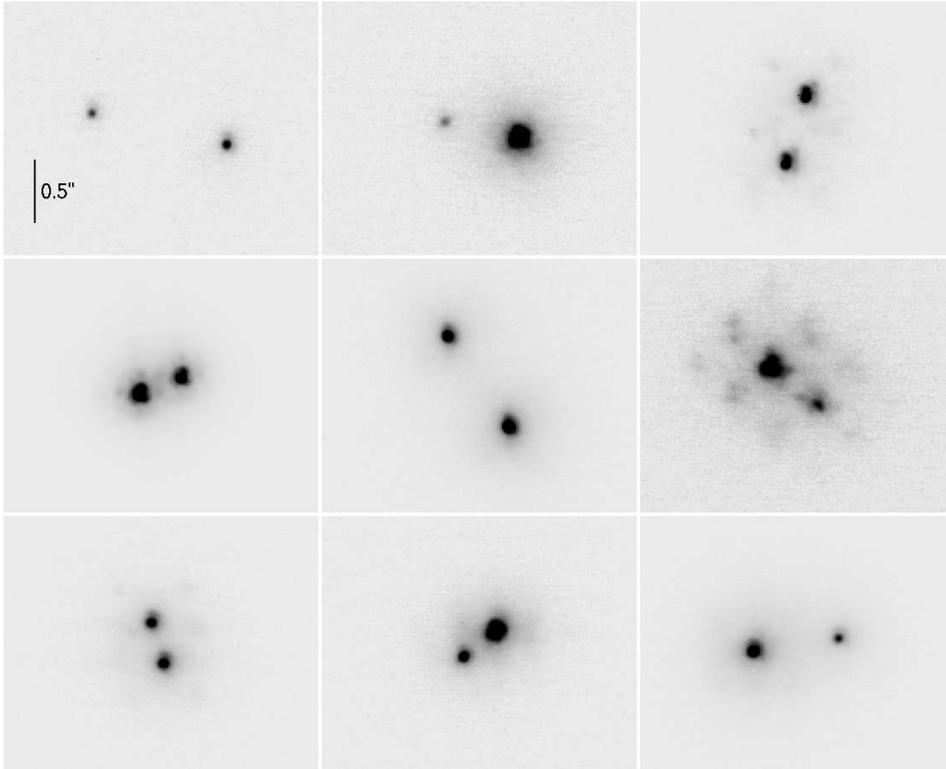}
\caption{\label{fig1}NACO H-band images of 9 of our 22 binary targets in the ONC.}
\end{figure}

Due to their close separations, all target binaries were observed with the help of Adaptive Optics (AO). We imaged all targets with VLT/NACO in $J$ and $H$ bands (see Fig.~\ref{fig1}) which will be combined with data from the literature to provide full $JHK$ photometry of all separate binary components in the sample. Furthermore, AO assisted spectroscopy of all targets were taken with NACO (R$\sim$1400, 16 sources) and Gemini/NIFS (R$\sim$5000, 6 sources) providing separate K-band spectra of all binary components.

\subsection{Reduction}
All spectra and images were reduced and extracted with custom {\em IDL} and {\em IRAF} procedures. Telluric correction was achieved through division by telluric standard stars which were observed close in time and at similar airmass as each target. All telluric standards are of spectral type B0--B9. Hence, in order to preserve the intrinsic Br$\gamma$ information from the target components, the telluric standard spectra had to be cleaned from the significant Br$\gamma$ absorption feature. This was achieved by fitting and dividing a Moffat line model to the absorption feature and thus removing the Br$\gamma$ line before division.

\begin{figure}[t]
\plotone{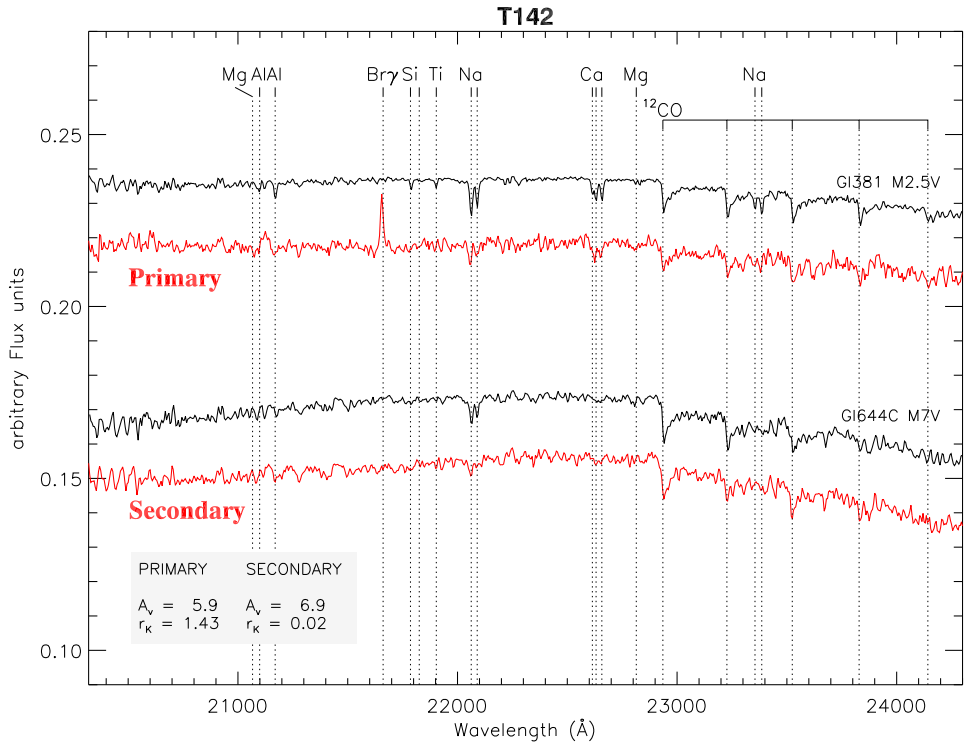}
\caption{\label{fig2}In red: NACO Spectra of both components of one of our target binaries. In black: artificially extinced and veiled template spectra \citep{ray09,cus05} to find spectral types, extinction A$_\mathrm{V}$, and veiling r$_\mathrm{K}$. The best fit spectral types are noted on the right and best A$_\mathrm{V}$ and r$_\mathrm{K}$ are reported in the box in the lower left.}
\end{figure}

In order to derive spectral type, extinction, and veiling for each component, the spectra were matched with template spectra from the IRTF Spectral Library \citep{ray09,cus05}. Best estimates were found from a 3-parameter least $\chi^2$ fit (Fig.~\ref{fig2}).

\section{First Results: Brackett Gamma Emission Statistics}
In the following, we will discuss only the spectroscopy results from a subset of the 22 observed targets. The reduced dataset consists of 16 fully reduced NACO spectra and will be amended by the rest of the targets and observational modes as soon as the data are coherently reduced and extracted.

Fig.~\ref{fig3} shows a census of Br$\gamma$ emission, as an indicator of ongoing accretion, observed in the components that were targeted with NACO. Components with significant Br$\gamma$ emission (peak to noise ratio $>$3) were considered accreting. Typical measured equivalent widths of successful detections were in the range of 0.5 to 5\,\AA, although also one strong emitter was detected with W(Br$\gamma$)=18.5$\pm$2.6\,\AA.

We see a clear preference for systems with non-accreting components (9 out of 16 binaries). Mixed pairs with one accreting component as well as systems with both components accreting are relatively rare. From these statistics we can draw the following conclusions:

\begin{figure}[t]
\plotone{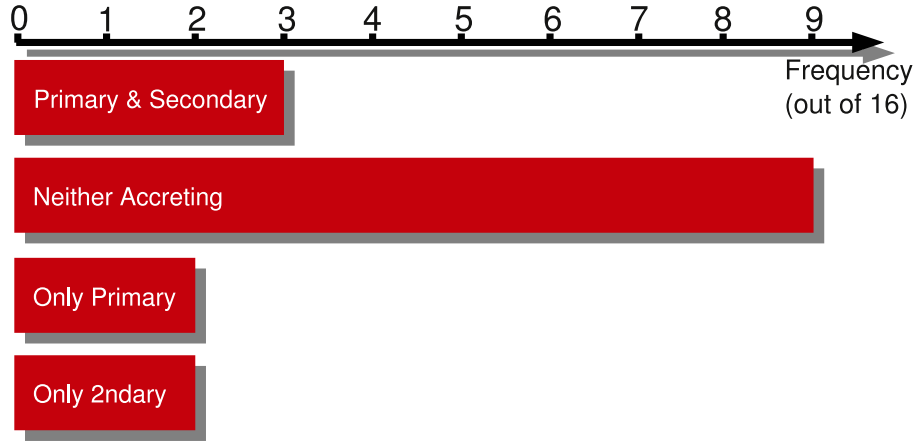}
\caption{\label{fig3}Accretion activity, as assessed through Br$\gamma$ emission, in the primary and/or secondary component of the 16 binaries observed with NACO spectroscopy. A thorough discussion of the detection limits is presented in Daemgen et al. (2011, {\em in prep.}).}
\end{figure}

\subsection{Accretion disks exist in primaries and/or secondaries.}
As was observed in other studies of various star-forming regions \citep[e.g.][]{mon07,pra03}, we find all combinations of accreting and non-accreting primary and secondary components. Furthermore, we observe a preference towards neither the higher mass (typically the primary) nor the lower mass component to be more likely to exhibit an accretion disk.

\subsection{ONC binaries are less frequently accreting than single stars.}
We see that most of the binary components do not show significant signs of accretion. Assuming that, as a consequence of the star formation process, initially all target components were in a state of accretion, most of the components---preferably both components of a binary---stop accreting within the short period of the lifetime of the ONC ,i.e.\ $\sim$1\,Myr. The statistics of Br$\gamma$ emission in the components (10 out of 32) would imply an accretion disk frequency of only $\sim$31$\pm$10\%. When compared to the disk frequency of single systems in the ONC \citep[55\%--90\%][]{hil98} it seems that a binary companion forces accretion to stop earlier than it would in a single system. 

We caution that the above frequency of binary accretion disks cannot directly be compared to the single star disk frequency, since the latter was measured by means of infrared excess and not accretion. The 31\% accretion disk frequency calculated above can, however, serve as a lower limit for the disk frequency in binaries: \citet{fed10} show that accretion disks in star forming regions are less frequently observed than the infrared excess (indicating hot circumstellar dust) of stars in the same regions. For associations younger than 5\,Myr, the accretion disk frequency is on average 9\% lower than the disk frequency detected by infrared excess. Additionally, Br$\gamma$ emission through accretion is considerably fainter than the simultaneously produced H$\alpha$ emission, the standard accretion measure also used by \citet{fed10}. Hence, objects might exist in the sample which would be detected as accreting by means of H$\alpha$ but not with Br$\gamma$ if the equivalent width is buried in the noise. The equivalent width of Br$\gamma$ can be estimated to be about $\sim$$1/4$ of the H$\alpha$ value (compare \citealt{edw94} and \citealt{naj96}). Hence, an H$\alpha$ equivalent width of 10\,\AA\ implies $W($Br$\gamma)\approx2.5$\,\AA. This is within the detection limits of most of our target components. Accordingly, although accreting targets might be missed by our Br$\gamma$ measurement, their number should be small.

Taking into account these considerations, the data would still point to an underrepresentation of accretion disks in binaries when comparing their frequency (31$\pm$10\%) to the first order estimation of the accretion disk fraction of single stars in the ONC ($\equiv$dust excess frequency reduced by 9\%, i.e.\ 46\%--81\%).

\subsection{\label{sec4.3}Binarity {\em does} influence disk evolution even for separations $>$100\,AU.}
Our ONC sample consists of binaries with separations of 100--400\,AU. Models \citep{may05} and results from other star-forming regions \citep[Ophiuchus, Taurus;][]{duc10} typically predict significant differences between single star evolution and binaries only for close systems $<$100\,AU. However, the statistics in Fig.~\ref{fig3} is not compatible with being composed from unrelated, randomly paired sources: assuming a disk frequency of 31\%, we statistically expect to find $\sim$1.5 sources with both components accreting and $\sim$7 sources with one component accreting. Despite the low number statistics, this seems to be not observed in our sample, but pairs of non-accreting components appear to be preferred.

This finding, in addition to the overall low disk frequency, strongly supports an impact of binarity on disk evolution even in systems as wide as 100--400\,AU.

\section{Outlook}
This is the {\em largest and most complete spatially resolved spectroscopic study of sub-arcsecond Orion binaries} to date. The assessment of spatially resolved $JHK$ photometry and NIR spectroscopy will allow us to infer not only the presence of disk accretion as described above, but also hot circumstellar dust as well as spectral type, mass, age, and luminosity of each individual component. The complete dataset will diagnose the differential evolution of disks in ONC binaries, complement multi-region studies, and provide input to binary disk evolution and planet formation theories.

\bibliography{daemgen_s}

\end{document}